\documentclass[twocolumn]{aastex631}

\newcommand\aastex{AAS\TeX}
\newcommand\latex{La\TeX}

\usepackage{amsmath}

\newcommand{\figsize}{0.45}

\newcommand{\xmm}{{\it XMM-Newton}}
\newcommand{\swift}{{\it Swift}-XRT}

\begin{document}

\title{Utilizing Maximum Variability to Discern TDE Emission from AGN Flares}

\correspondingauthor{Samaresh Mondal}
\email{samaresh.astro@gmail.com}

\author[0000-0001-9042-965X]{Samaresh Mondal}
\affiliation{Department of Astronomy, University of Illinois, 1002 W. Green St., Urbana, IL 61801, USA}

\author[0000-0002-4235-7337]{K. Decker French}
\affiliation{Department of Astronomy, University of Illinois, 1002 W. Green St., Urbana, IL 61801, USA}





\begin{abstract}
X-ray emission arising from active galactic nucleus (AGN) activity may potentially mimic the expected emission of tidal disruption events (TDEs). Ongoing and upcoming wide-field X-ray surveys will detect thousands of TDE-like sources, and classifying them securely as TDEs or AGNs is a challenging task. To this aim, we measure the average X-ray variability of AGNs and derive a threshold of maximum variation as a function of time separating the TDEs from AGN flares. For the comparison between TDE and AGN X-ray variability, we cross-match the publicly available \xmm\ and \swift\ point source catalogs with the Million Quasars Catalog and optically selected TDEs. Then we compute the X-ray structure function (SF) and maximum variability of the AGN and TDE samples. The X-ray SF of AGNs has a power-law index $\gamma\sim0.11-0.14$ when fitted with a simple power-law model. However, the SF of AGNs is best described by a broken power-law or a power exponential model with a damping time scale $\tau=950\pm300$ days. The maximum variability comparison between TDE and simulated AGN light curves indicates they have a similar order of variation on a time scale of less than 20 days. However, at a longer time scale of $\sim20$ days or more, the large-scale variations expected from power-law-like decay in TDEs become less frequent in AGNs. Furthermore, we compare the maximum variability of eROSITA TDE candidates with AGN, finding that many of the eROSITA-DE TDE candidates are consistent with flares from AGNs, and may not be TDEs.          
\end{abstract}



\section{Introduction} \label{sec:intro}
Tidal disruption events (TDEs) are astronomical phenomena that occur when a star ventures too close to a supermassive black hole (SMBH) and is torn apart by the black hole's intense gravitational tidal forces \citep[see][for a review]{gezari2021}. TDEs were proposed as a theoretical concept in the 1970s \citep{hills1975,lidskii1979}, and they occur when the peri-center of a star's orbit passes through the tidal disruption radius of an SMBH. The tidal radius depends on the mass and radius of the orbiting star ($M_*$, $R_*$) as well as the mass of the SMBH ($M_{\rm BH}$), and is given by $R_{\rm T}\approx R_*(\frac{M_{\rm BH}}{M_*})^{\frac{1}{3}}$. For SMBHs with masses larger than $M_{\rm crit}\approx10^8(\frac{R_*}{R_{\odot}})^\frac{3}{2}(\frac{M_*}{M_{\odot}})^{-\frac{1}{2}}\ M_{\odot}$, the tidal radius is found inside the event horizon of the SMBH. In this case, the star becomes trapped within the event horizon before being disrupted, and no flare is observed. When a star passes through the tidal radius of an SMBH, the tidal force overcomes the star's binding energy. Here, roughly half of the stellar debris remains bound to the SMBH, and the rest is ejected at escape velocity \citep{rees1988}. The bound material forms an accretion disk around the SMBH, and the accretion process produces a bright, transient electromagnetic flare that can last for weeks to months. The disrupted star's material begins to accrete onto the black hole, which causes the X-ray luminosity to rise to or even exceed the Eddington luminosity on a time scale of days to weeks. The brightness then gradually decays over months to years, typically following a power-law decay with time.

TDEs provide a unique opportunity to study the properties and behavior of SMBHs that allow astronomers to detect the presence of SMBHs in distant galaxies. The first TDEs were discovered in 1990 by the \emph{ROSAT} X-ray satellite, which detected a dramatic and sudden X-ray flare from the galaxies NGC 5905, RXJ1242--1119, RXJ1624+7554, RXJ1420+5334 \citep{bade1996,komossa1999a,komossa1999b,grupe1999,greiner2000}. In optical wavelengths, roughly 100 TDEs have been discovered by various surveys \citep{vanVelzen2021,hammerstein2023,yao2023}. 
Among those, around $\sim70$ are detected in X-rays \citep{goldtooth2023,guolo2024}. The origin of optical emission from TDEs is not yet clear. Several possibilities for the optical emission have been proposed, such as reprocessing of X-ray emission from the accretion disk by optically thick material surrounding the disk \citep{guillochon2013,roth2016,dai2018} or from shocks in the outer part of the disk as the debris streams collide with one another \citep{piran2015}.

Due to the lack of good sampling, the shape of TDE light curves may resemble other transients, such as supernovae and flares from active galactic nucleus (AGN). TDEs can be easily distinguished from supernovae by post-peak light curves \citep{muthukrishna2019}. However, the separation of TDEs from AGN can be more difficult as AGN variability is stochastic in nature and the variation in flux can span several orders \citep{markowitz2004,rumbaugh2018}. AGNs are variable on time scales ranging from less than a day to decades, across all wavelengths. This variability can be caused by changes in the accretion rate, the hot corona, and the varying impact of dust obscuration. The magnitude of variability tends to increase with frequency, such that X-ray variability is a common feature of AGN \citep{mushotzky1993}. On the other hand, TDE variability is largely driven by changes in the mass capturing rate at the tidal radius, which shows a sharp rise and monotonic power-law decay over a long time scale \citep{rees1988}. If an AGN is caught during the flaring phase, it can be easily misclassified as a TDE. eROSITA has already discovered some AGNs during flare episodes \citep{saha2023,krishnan2024}, and without spectroscopic information (X-ray or optical), it can be extremely difficult to distinguish between AGNs and TDEs. With the advancement of wide-field all-sky optical/X-ray monitors, an abundance of TDEs, as well as AGNs at the flaring phase, will be discovered. Furthermore, a vast majority of them are likely to have no spectroscopic or prior classification. Hence, there is a need to characterize these sources only from the variability in light curves. Methods for identifying TDEs using modest increases in the X-ray flux may be contaminated by AGN flares, and here we aim to quantify this contamination rate.

Using a sample of four TDEs and four extremely variable AGN, \citet{auchettl2018} demonstrated that AGN flares can have amplitudes and durations similar to TDEs, but TDEs decay significantly more monotonically, and their X-ray emission exhibits little variation in spectral hardness as a function of time. These results indicate that separation of TDEs and AGN in X-ray light curves can be achieved, however, the sample of four TDEs used by \citet{auchettl2018} is small and limited to cases where prompt X-ray emission is seen after the discovery. Recent observations of X-ray light curves of TDEs identified in optical wavelengths show a broad range in behavior, with many showing delayed X-ray flares \citep{gezari2017, guolo2024}. 

It is important to quantify the rate of AGN flare contamination in TDE searches, as even rare AGN flares can have an impact due to the low rate of TDEs. Studies of quasar variability show that less than 10\% of quasars show maximum variation over one magnitude on a time scale of years to decades \citep{macleod2016, graham2017, rumbaugh2018}. On the other hand, at a time scale of a few weeks to months, the maximum variation of TDE is expected to be much larger than AGNs. In this work, we aim to quantify the rate of AGN flares in the X-ray that may contaminate TDE searches, using a large sample of TDEs and AGNs. 

AGN variability in the optical and X-ray has been measured using a structure function (SF) to characterize the ensemble-averaged variability as a function of time scale. \citet{vagnetti2011} used a sample of 412 \xmm\ AGNs and found that the SF can be best described by a power-law model with an index of $0.1\pm0.01$. Later, \citet{vagnetti2016} extended the analysis of SF for a sample of 2112 AGNs by cross-matching \xmm\ and the SDSS catalog and found that the SF has a power-law index of $\sim0.12$ independent of AGN luminosity; however, the normalization of the SF inversely depends on AGN luminosity. \citet{middei2017} obtained the SF from a sample of 2700 \xmm\ AGNs as well as 281 AGNs that have both \xmm\ and {\it ROSAT} \citep{voges1999} observations. They found that the SF continues to increase up to a rest-frame time scale of twenty years with a power-law index of $0.15\pm0.01$. Recently \citet{georgakakis2024} obtained the SF from a sample of $\sim9000$ AGN that have been observed by both \xmm\ and eROSITA \citep{predehl2021} and obtained the SF power-law index of $0.12\pm0.01$. These studies suggest a monotonically increasing trend of the SF in the decadal rest-frame time intervals with logarithmic slopes of 0.1--0.15, with no turnover yet seen at large time scales.

However, at longer time intervals, the AGN SF is expected to flatten to a constant value. The flattening at a large time scale is due to the thermal time scale of the outer part of the accretion disk. Optical studies of AGN variability show a flattening of SF at a rest-frame time scale of $\sim350$ days \citep{kelly2009,macleod2010}. In addition to that, the flattening time scale and the amplitude of variability seem to correlate and anti-correlate with the BH mass, respectively \citep{kelly2009,macleod2010,kozlowski2016,burke2021}. The flattening time scale and the amplitude of variability have an important consequence in estimating the rate of AGN flares, as the former will determine how often the large-scale variations are seen, whereas the latter determines how large the flux variations are. Thus, an accurate X-ray structure function for AGN is needed in order to predict the rate of TDE-like flares.

The paper is organized as follows: we discuss our archival data from \xmm\ and \swift\ and introduce a parameter for the maximum variability (MaxVar) as a function of time scale in \S2. In \S3 we measure the SF to determine AGN variability and MaxVar for the AGN and TDE samples, and use the MaxVar function to quantify the rate of TDE-like flares that would be observed in AGN. In \S4 we discuss the shape of the AGN SF and compare the MaxVar of mock AGN light curves and eROSITA TDE candidates. We conclude in \S5.

\section{Methods}
\subsection{Data Set Preparation} \label{sec:data}
To compile the \xmm\ \citep{jansen2001} long-term light curve of point sources, we utilize the \xmm\ stack catalog 4XMM-DR14s \citep{traulsen2019,traulsen2020}. We only select sources with more than one detection using the flag \texttt{N\_CONTRIB\footnote{Indicating how many observations available for each source.}$>$1}, with the maximum likelihood of each detection being \texttt{EP\_DET\_ML\footnote{The maximum likelihood is obtained from the point spread function (PSF) modeling of the X-ray sources.}$>$14}. Sources with a maximum likelihood of $>$14 ensure a detection significance above 5$\sigma$ \citep{webb2020}. We use the selection criteria \texttt{EXTENT\footnote{Describes the source type, in case the likelihood of the source being extended falls below a threshold of four or the extent radius of the PSF model is below $6''$ then the source extent is set to zero, indicating a point source.}$=$0} indicating a point-like source. Furthermore, to avoid spurious and/or problematic cases of source detection, we chose \texttt{STACK\_FLAG\footnote{0 indicates detection without any warning, and 1 indicates reduced detection quality in at least one instrument.}$<$1}. This resulted in a total number of 21420 unique point sources with more than one \xmm\ detection.

For the construction of the \swift\ \citep{burrows2005} long-term light curve, we use the Living \swift\ Point Source (LSXPS) catalog \citep{evans2023}. We removed stacked detections by using the flag \texttt{IsStackedImage\footnote{Indicating whether detection was made in a single \swift\ observation or from stacking of multiple observations.}$=$0}, to avoid double counting. In addition, we only select sources that were detected in the total band 0.3--10 keV using the flag \texttt{Band\footnote{Indicate the energy band of the detections: 0=0.3--10 keV, 1=0.3--1 keV, 2=1--2 keV, 3=2--10 keV}$=$0} and also avoid spurious detection using the flag \texttt{DetFlag\footnote{Indicating the detection quality. 0, 1, and 2, meaning good, reasonable, and poor, respectively. Higher values indicate more warnings.}$<=$1}. This resulted in a total of 38773 unique point sources with more than one \swift\ detection.

\begin{figure}[]
\centering
\includegraphics[width=\figsize\textwidth]{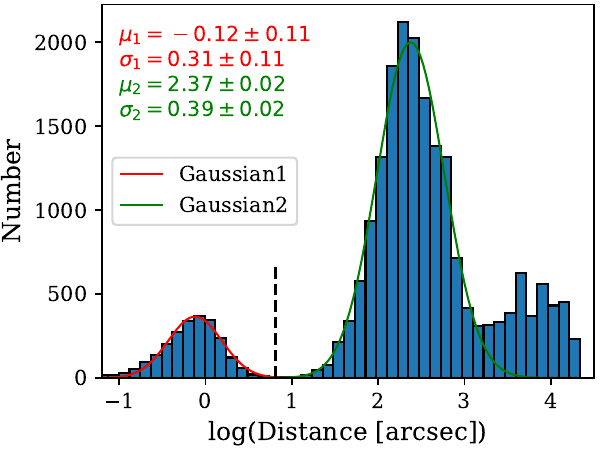}
\includegraphics[width=\figsize\textwidth]{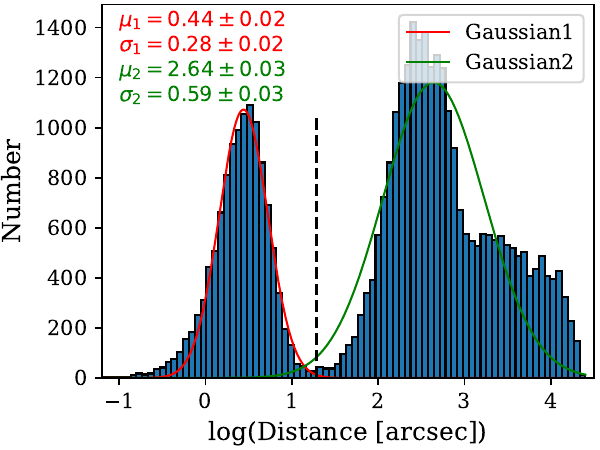}
\caption{The nearest neighbor distance distribution obtained by comparing the \xmm\ (top panel) and \swift\ (bottom panel) sources with MILLIQUAS catalog. The histograms show a bimodal distribution, which is modeled with two Gaussians. We use these fits to obtain the maximum distance used for cross-matching the X-ray sources with the AGN sample. The black dashed line shows the 3$\sigma$ distance from the mean of the nearest Gaussian.}
\label{fig:dist}
\end{figure}

To prepare the AGN sample, we cross-match the \xmm\ and \swift\ point sources with the Million Quasars (MILLIQUAS) Catalog \citep{flesch2023}. First, we compute the nearest-neighbor distance between the \xmm\ and \swift\ sources with sources in the MILLIQUAS catalog. Figure \ref{fig:dist} shows the distribution of nearest-neighbor distance in log scale obtained from the 21420 \xmm\ (top panel) and 38773 \swift\ (bottom panel) sources. The distribution shows two peaks, with the nearest one indicating the actual association of the X-ray sources with MILLIQUAS AGNs, while the sources belonging to the farthest peak indicate random association. We model the bimodal distribution of log distance with two Gaussians and use the 3$\sigma$ distance from the mean of the nearest Gaussian for cross-matching, which is indicated by the black dashed line in Figure \ref{fig:dist}. Therefore, we use a distance of $6''$ and $19''$ for cross-matching between the \xmm\ and \swift\ sources with the MILLIQUAS catalog, respectively. The cross-matching distances are well within the resolving power of both satellites. The PSF of \xmm\ EPIC-pn \citep{struder2001} and \swift\ have a half-power diameter (or full width at half maximum) of $15''$ ($6''$) and $18''$ ($7''$), respectively. For each source in the MILLIQUAS catalog, we check if there are two or more X-ray sources within the limiting distance for cross-matching; then, in such a case, we do not consider that source. Such a situation can arise in case there are X-ray binaries and/or ultra-luminous X-ray sources within $6''$ and $19''$ from the central AGN for \xmm\ and \swift\ data sets, respectively. After doing this filtering, we find 2363 AGNs in \xmm\ and 11649 AGNs in \swift\ catalogs with more than one detection with redshift measurements in the MILLIQUAS catalog. The redshift values in the MILLIQUAS catalog represent spectroscopic measurements that are taken from literature or obtained from catalogs released by Dark Energy Spectroscopic Instrument \citep{desi2024} and Sloan Digital Sky Survey-DR18 \citep{almeida2023}.

For the compilation of the TDE sample, we search for all the optical TDEs discovered through 2025-03-01 using the  TNS\footnote{https://www.wis-tns.org/} webpage. In addition to that, we include the $\sim70$ X-ray detected TDEs listed in \citet{goldtooth2023}. Our final list includes 213 TDEs, which we use for cross-matching with \xmm\ and \swift. We use the same separation distance as AGN ($6''$ for \xmm\ and $19''$ for \swift) for cross-matching with the TDEs and X-ray source list. The final X-ray detected TDE list includes 16 sources with more than one \xmm\ detection and 50 sources with more than one \swift\ detection. Among the TDEs that have more than one detection either in \xmm\ or \swift, 23 are not found in either \citet{goldtooth2023} or \citet{guolo2024}, indicating new TDE-X-ray associations. The majority of these new objects (16/23) are from 2022-2024, past the dates included in these studies. In contrast, the TDE sample in \citet{goldtooth2023} includes sources discovered through November 2020. \citet{guolo2024} includes TDEs discovered from 2014 to 2021.

\subsection{SF and MaxVar computation}\label{sec:sf_max_var}
We use SF and MaxVar as tools to compare the long-term variability of X-ray-selected TDEs and AGNs. 

SF is a powerful tool describing the mean change between two observations separated by a time lag $\Delta t$. A number of different mathematical formulations for SF have been used in the past \citep{simonetti1985,diclemente1996}; however, we used the following definition of the ensemble-averaged SF for X-ray light curves: 
\begin{equation}
    {\rm SF}(\Delta t)=\sqrt{\big<( {\rm log}f_{{\rm x},i} - {\rm log}f_{{\rm x},j} )^2\big>-\big<\sigma_{ij}^2\big>}
\end{equation}
where $\sigma_{ij}^2$=$\big(\frac{\Delta f_{{\rm x},i}}{f_{{\rm x},i}\times\ln(10)}\big)^2+\big(\frac{\Delta f_{{\rm x},j}}{f_{{\rm x},j}\times\ln(10)}\big)^2$ and $f_{\rm x}$, $\Delta f_{\rm x}$ are X-ray flux and its error in 0.3--10 keV band, respectively. For a given $\Delta t$, the mean is calculated for all possible combinations of $i$ and $j$ that satisfy $\Delta t=\frac{abs(t_i-t_j)}{1+z}$ across all sources in a given sample, where $i$ and $j$ are the time indices of the entries in the light curve and $z$ is the redshift of the source. We bin the SF in log space with a step size of $\Delta {\rm log}(\Delta t)=0.2$. For cases where a bin has $<20$ points, we increase the bin size to 0.4 or 0.6 dex to ensure that each bin has at least 20 data points. This requirement allows for the error in each bin not to be dominated by Poisson statistics in the low count regime, and to be approximated as Gaussian. For the uncertainties in each bin, we use the standard error of the mean.

\begin{figure}[]
\centering
\includegraphics[width=\figsize\textwidth]{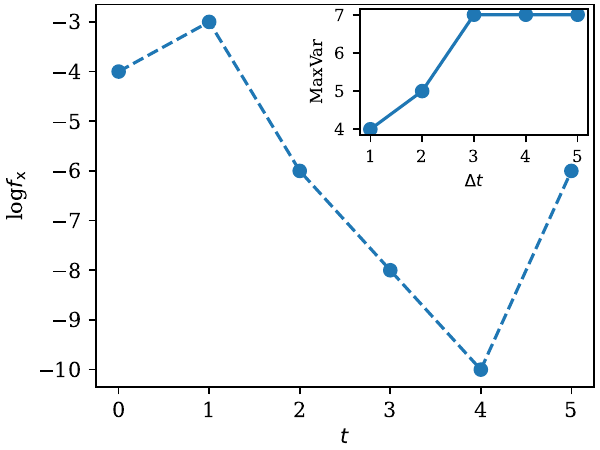}
\caption{The illustration of MaxVar from an artificial light curve that consists of six measurements.}
\label{fig:maxvar_example}
\end{figure}

Next, we compute the MaxVar using the formula given below,
\begin{equation}
    {\rm MaxVar}(\Delta t)=\big< max\big(max({\rm log}f_{\rm x}) - min({\rm log}f_{\rm x})\big) \big>
    \label{eq:eq_maxvar}
\end{equation}
where the $max$ and $min$ are maximum and minimum of ${\rm log}f_{\rm x}$ computed for a single source individually for all possible $\Delta t$ combinations. Then we take the maximum of the maximum difference of ${\rm log}f_{\rm x}$ in a single source for all possible unique $\Delta t$. In the end, the mean is computed for each unique $\Delta t$ across different sources in the given sample. In Figure \ref{fig:maxvar_example}, we illustrate the MaxVar computed using Equation \ref{eq:eq_maxvar} for an artificial light curve consisting of six measurements. In case there is a single erroneous detection with huge variation/deviation from the average flux, the MaxVar on a longer time scale will be driven by this single detection. Such a situation can arise in case the exposure is too low, resulting in a detection with a large error bar or observations affected by soft proton flares. To avoid such a situation, before computing the MaxVar, we use sigma clipping to remove any single variation that exceeds the 5$\sigma$ limit. The sigma clipping is only activated when there is a single detection $>5\sigma$ from the mean, and not when there are more detections above the threshold. The sigma clipping does not have any significant effect on the MaxVar computation, as it does not remove the long-time scale flares that we are most interested in. The sigma clipping lowers the MaxVar only by 2.8\% and 8.5\% at a shorter time scale of $\sim10$ days and a longer time scale of $\sim1000$ days, respectively.

Previous work on the maximum variability typical of AGN has focused on identifying AGN with extreme variability changes, rather than determining rates of these events, and has typically used optical instead of X-ray data. \citet{macleod2016} and \citet{rumbaugh2018} consider quasars where the maximum $g$ band change is at least one magnitude\footnote{Note that these optical studies are considering variation in units of magnitude, where a 1 mag variation will indicate a flux variation of $10^{1/2.5}$. On the other hand, in our case, MaxVar has units of dex; therefore, a $\rm MaxVar=1$ dex indicates a variation of one order (10$\times$) in flux.}. In related work, \citet{graham2017} quantified the extreme end of AGN variability by identifying flares in excess of the typical variability as a function of time scale. Our definition of MaxVar is driven by the use of X-ray observations and the need to measure the rates of these extreme flaring events.

\section{Results}\label{sec:results}

\subsection{SF Analysis}

\begin{figure}[]
\centering
\includegraphics[width=\figsize\textwidth]{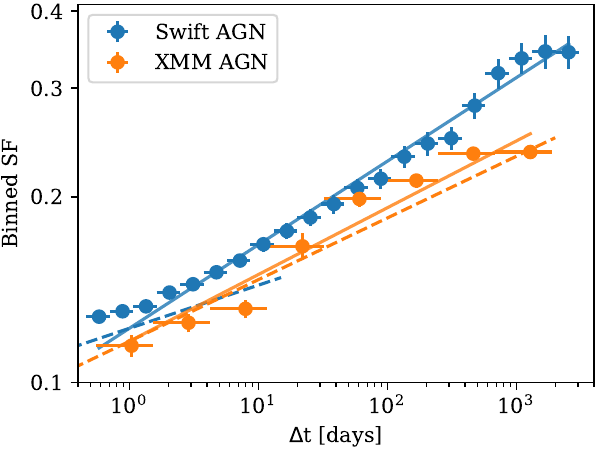}
\caption{SF (in log-log scale) of AGN sample computed from the \xmm\ and \swift\ data sets. The continuous lines indicate a simple power-law model fit to the data. The dashed lines indicate a slightly different power-law index obtained in a similar study by \citet{vagnetti2011} using \xmm\ and \swift\ data sets.}
\label{fig:sf_agn}
\end{figure}

\begin{deluxetable*}{c c c c c c c}
\label{tab:sf_agn_table}
\tablecaption{Best fit parameters of the fitted models to the SF of \xmm\ and \swift\ AGN sample. $A$ is the normalization of the simple power-law and power-law+const models at $\Delta t=1$ day. $\rm SF_0$ and $\rm SF_{\infty}$ are the normalizations of the broken power-law and power exponential models at the time scale $\Delta t_0$ and infinity, respectively. The p-value of the power-law+const model is computed by comparing the $\chi^2$ and $d.o.f$ with the simple power-law model, and the p-value of the broken power-law or power exponential models is computed by comparing the $\chi^2$ and $d.o.f$ with the power-law+const model.}
\tablehead{
   \colhead{Model} & \colhead{$\gamma /\gamma_1/ \beta$} & \colhead{$\Delta t_0$/$\tau$} & \colhead{$A$/$\rm SF_{0}$/$\rm SF_{\infty}$} & \colhead{const} & \colhead{$\chi^2/d.o.f$} & \colhead{p-value}
}
\startdata
        \multicolumn{7}{c}{\swift\ AGN SF} \\ \hline
        power-law & $0.135\pm0.004$ & & $0.122\pm0.003$ & & 74/19\\ \hline
        power-law+const & $0.20\pm0.03$ & & $0.06\pm0.02$ & $0.07\pm0.02$ & 15/18 & $1.9\times10^{-7}$\\ \hline
        broken power-law+const & $0.28\pm0.02$ & $1281\pm104$ & $0.244\pm0.008$ & $0.099\pm0.007$ & $7/17$ & $3.8\times10^{-4}$\\ \hline
        power exponential+const & $0.73\pm0.09$ & $930\pm328$ & $0.26\pm0.03$ & $0.113\pm0.008$ & $13/17$ & $1.2\times10^{-1}$\\ \hline
        \multicolumn{7}{c}{\xmm\ AGN SF} \\ \hline
        power-law & $0.11\pm0.01$ & & $0.114\pm0.006$ & & 34/6\\ \hline
        power-law+const & $0.12\pm0.11$ & & $0.1\pm0.15$ & $0.01\pm0.15$ & 34/5 & $1$\\ \hline
        broken power-law+const & $0.23\pm0.01$ & $273\pm98$ & $0.17\pm0.05$ & $0.06\pm0.05$ & $9/4$ & $2.9\times10^{-2}$\\ \hline
        power exponential+const & $0.8\pm0.2$ & $989\pm256$ & $0.15\pm0.02$ & $0.09\pm0.01$ & $7/4$ & $1.7\times10^{-2}$\\
\enddata
\end{deluxetable*}

Figure \ref{fig:sf_agn} shows the SF of the AGN sample for the \xmm\ and \swift\ data sets. The AGN SF in both data sets shows a linearly increasing trend with $\Delta t$ in log-log space. However, the slope of the AGN SF obtained from \xmm\ data sets is slightly flatter than \swift\ AGNs. A simple power-law model of ${\rm SF}(\Delta t)=A(\Delta t)^{\gamma}$  fit to the data results in slope $\gamma=0.11\pm0.01,0.135\pm0.004$ and the normalization $A=0.114\pm0.006,0.122\pm0.003$, for \xmm\ and \swift\ data sets, respectively. These power-law slopes broadly agree with the range of 0.1--0.15 seen in previous studies \citep{vagnetti2011,vagnetti2016,middei2017,georgakakis2024}. In Figure \ref{fig:sf_agn}, we show for comparison the best-fit power-law models from \citet{vagnetti2011}, although we caution that the \swift\ sample used in our analysis contains a larger number of AGN with observations over a longer time scale. In contrast with previous work, we see evidence for a turnover at long time scales, which motivates the use of models incorporating this turnover.

\begin{figure}[]
\centering
\includegraphics[width=\figsize\textwidth]{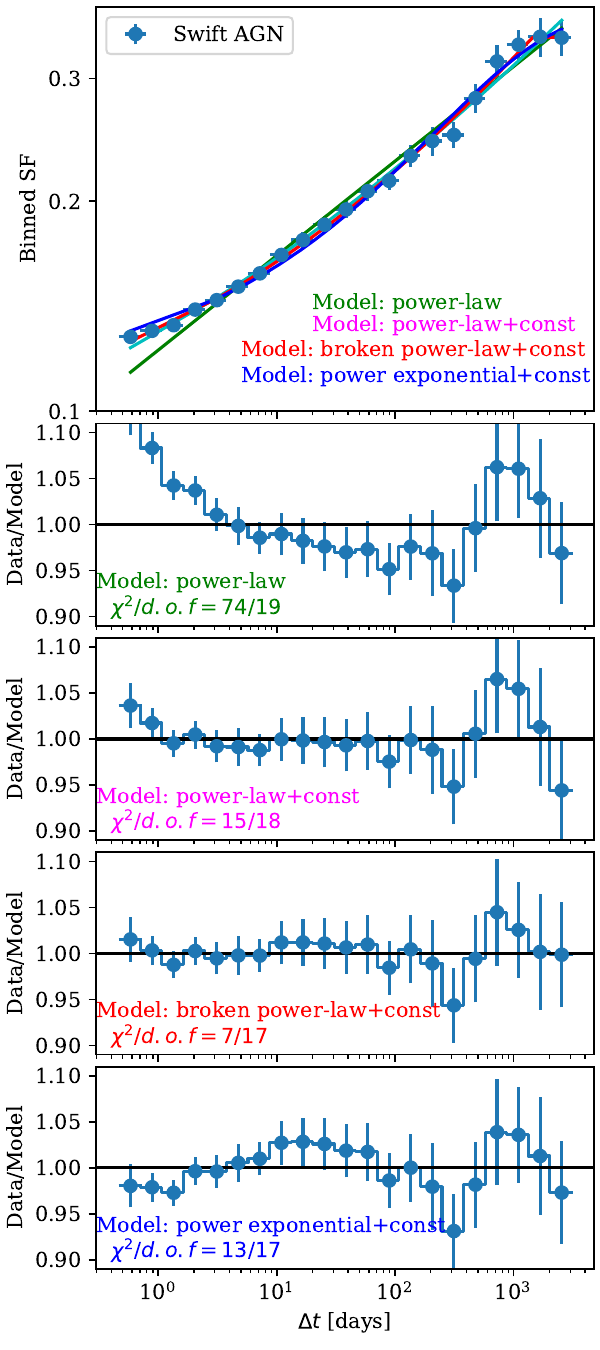}
\caption{Comparison of different models fit to \swift\ AGN SF (top panel in log-log scale). A broken power-law model provides a better fit compared to a simple power-law model, which is visible in the data/model plot.}
\label{fig:swift_sf_fits}
\end{figure}

The SF of AGN is expected to have a break above which the variability becomes nearly constant. To test the presence of a break, we fit the SF using a broken power-law model in which the power-law index is $\gamma_1$ below the break time scale $\Delta t_0$ and zero above the break time scale. Even though we subtracted the noise term while computing the SF, that should take into account the constant level of variability expected from the measurement error. A small amount of excess variability is still present at a time scale $\Delta t$ below one day (see the second panel of Figure \ref{fig:swift_sf_fits}). The constant term improves the fit by $\Delta \chi^2=59$ when compared to the simple power-law model (third panel of Figure \ref{fig:swift_sf_fits}). Therefore, for the subsequent modeling, we add a constant term to take into account the excess variability at a very low time scale. The broken power-law model is defined as,
\begin{equation}
    {\rm SF}(\Delta t) = 
    \begin{cases}
      {\rm SF_0}(\frac{\Delta t}{\Delta t_0})^{\gamma_1}+{\rm const}, & \text{if}\ \Delta t<\Delta t_0 \\
      {\rm SF_0}+{\rm const}, & \text{if}\ \Delta t>=\Delta t_0
    \end{cases}
\end{equation}
where SF$_0$ is the SF at the break time scale $\Delta t_0$. Fitting with this model significantly improves the fit compared to the power-law+const model. There is a $\Delta \chi^2=8$ improvement for one additional degrees of freedom ($d.o.f$). We did an F-test of the $\chi^2$ improvement, which gives a p-$\rm value=3.8\times10^{-4}$ that suggests the power-law+const model can be rejected over the broken power-law+const model at more than $3\sigma$ confidence level. The residuals of the fitted model are shown in the fourth panel of Figure \ref{fig:swift_sf_fits}. The best-fit parameters with $1\sigma$ uncertainty are given in Table \ref{tab:sf_agn_table}.

The variability in AGNs can also be described using a power exponential covariance matrix for the signal \citep{kelly2009,macleod2010,kozlowski2010}. Such a power exponential covariance matrix will generate an SF of the form $\rm SF(\Delta t)=SF_{\infty}\sqrt{1-exp\big(-(\frac{\Delta t}{\tau})^\beta\big)}$, where $\tau$ is the damping time scale and $\rm SF_{\infty}$ is the SF at infinity. Therefore, we also fitted the data with the power exponential model plus a constant. While doing the fit using this model, we kept $\beta$ as a free parameter. The optical light curve of AGNs is best described by a damped random walk model ($\beta=1$) that gives an SF power-law index of $\gamma\sim0.5$ below the damping time scale $\tau$. As in our case, the shape of the SF is much flatter than the SF seen in optical light curves; therefore, we kept the $\beta$ as free. Fitting with this model provides a marginal improvement over the power-law+const model by $\Delta \chi^2=2$ for one additional $d.o.f$. The best-fit parameters are given in Table \ref{tab:sf_agn_table}, and the residuals are shown in the fifth panel of Figure \ref{fig:swift_sf_fits}. Later, we used the best-fit parameters of this model to simulate light curves.

\begin{figure}[]
\centering
\includegraphics[width=\figsize\textwidth]{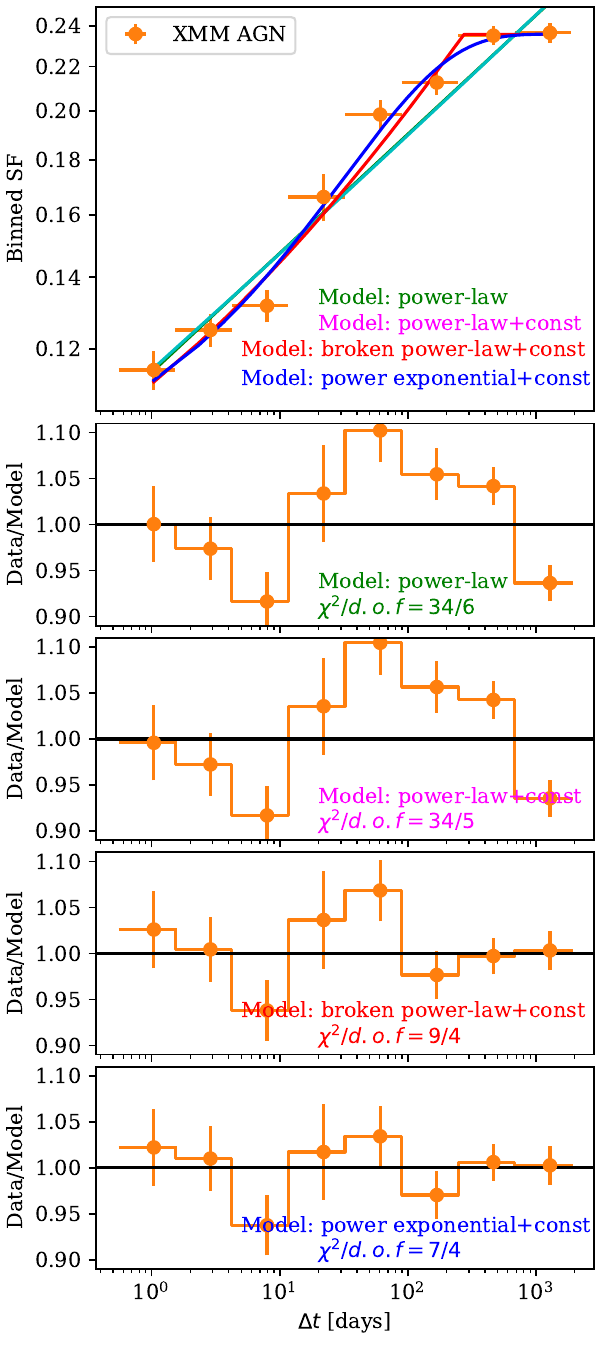}
\caption{Comparison of different models fit to AGN SF (top panel in log-log scale) for \xmm\ data sets. There is a significant improvement in the fit after incorporating a broken power-law model over a simple power-law model, which can be seen in the data/model plot.}
\label{fig:xmm_sf_fits}
\end{figure}

We also compared various model fits to the AGN SF obtained from the \xmm\ data set, and the best-fit parameters are given in Table \ref{tab:sf_agn_table}. The broken power-law+const model provides a better fit compared to the power-law+const model with $\Delta \chi^2=25$ improvement for one extra $d.o.f$. The F-test indicates the power-law+const model can be rejected over the broken power-law+const model at $>2\sigma$ confidence level. The comparison of the two model fits is shown in Figure \ref{fig:xmm_sf_fits}. Furthermore, we fit the AGN SF of \xmm\ data with a power exponential form that provides $\Delta\chi^2=27$ improvement for one additional $d.o.f$ compared to the power-law+const model. We computed the F-test that gives a p-value of 0.017, suggesting the power exponential model is preferred over the power-law+const model at $>2.5\sigma$ confidence level. The residuals of the fitted model are shown in the fifth panel of Figure \ref{fig:xmm_sf_fits} (further details are discussed in \S4.1).

\subsection{MaxVar Analysis}

\begin{figure}[]
\centering
\includegraphics[width=\figsize\textwidth]{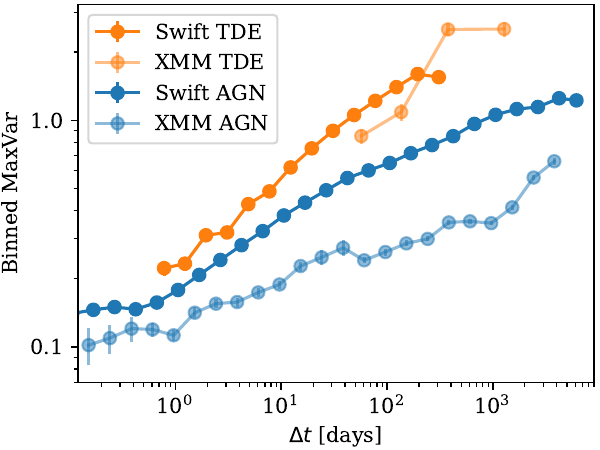}
\caption{The comparison of MaxVar between AGNs and TDEs for the different data sets. At a longer time scale, the MaxVar of TDEs is much higher than AGNs due to the fact that TDEs follow a monotonic decline.}
\label{fig:maxvar}
\end{figure}

To quantify the time scale at which the variability of TDEs dominates over the AGNs, we compute the MaxVar using Equation \ref{eq:eq_maxvar}. Figure \ref{fig:maxvar} shows the MaxVar of AGNs and TDEs for \xmm\ and \swift\ data sets. The MaxVar of any variable source is expected to increase with $\Delta t$.
The MaxVar of the TDEs is much higher compared to AGNs at longer time scales, for both the \xmm\ and \swift\ data sets. The MaxVar of \xmm\ TDEs is limited to $\Delta t>20$ days due to the lack of short cadence observations. However, at shorter time scales, the separation between the MaxVar of TDEs and AGNs decreases and may become comparable. 

We simulate AGN light curves that are representative of \xmm\ and \swift\ data sets to estimate the rate of AGN flares that have a similar or higher magnitude of variation than TDEs. The simulated AGN light curves have similar variability in their time series, by construction, as the actual \xmm\ and \swift\ light curves on a time scale of 5000 days.

\begin{figure}[]
\centering
\includegraphics[width=\figsize\textwidth]{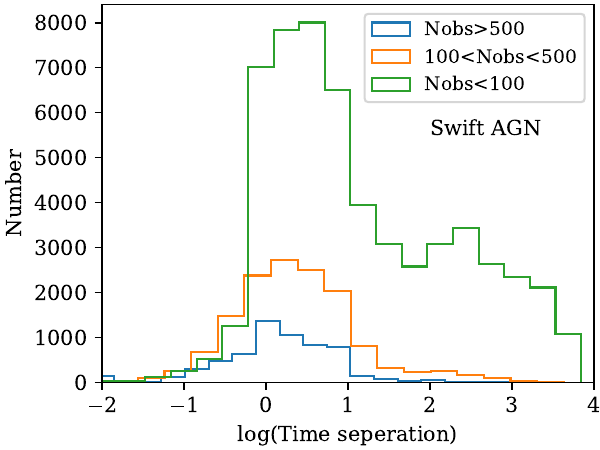}
\includegraphics[width=\figsize\textwidth]{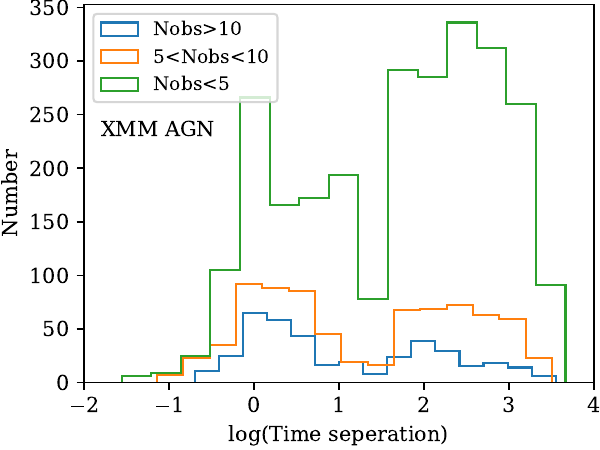}
\caption{The distribution of sampling time in days for \swift\ and \xmm\ AGN light curves. The x-axis represents the time difference between two consecutive measurements in the light curves. The colors represent light curves with varying numbers of observation points.}
\label{fig:dt_sampling}
\end{figure}

The MaxVar at short time scales is significantly affected by the gaps in observational data sets. The gaps in the data set decrease the MaxVar than the value expected from well-sampled light curves. The irregular observations decrease the probability of precisely catching the flares and dips, hence reducing the MaxVar. Therefore, to better quantify the effects of observational gaps in the light curve, we simulated AGN light curves using the parameters $\rm SF_{\infty}$, $\tau$, and $\beta$ obtained from fitting the SF using a power exponential model. The AGN variability is stochastic, and a power exponential covariance matrix best describes the light curves. We use the following covariance matrix for the signal 
\begin{equation}
    {\rm cov}(\Delta t)=\frac{\rm SF_{\infty}^2}{2}e^{(-\frac{\Delta t}{\tau})^\beta}
\end{equation}
where $\rm SF_{\infty}=0.26$, $\tau=930$ days and $\beta=0.73$ for the simulation of \swift\ AGN light curves and $\rm SF_{\infty}=0.15$, $\tau=990$ days and $\beta=0.8$ for \xmm\ AGN light curves. We simulated 2000 \xmm\ and \swift\ light curves with a cadence of 5 days and a duration of 5000 days. The typical sampling times in \swift\ \xmm\ light curves are in between 1--10 days as indicated by the first peak in the Figure \ref{fig:dt_sampling}, therefore we chose a 5-day cadence for the simulated light curves. For the simulation, the mean count rate of each light curve is drawn from a Gaussian distribution representing the actual distribution of the count rate in \xmm\ and \swift\ light curves. Furthermore, we add a noise term in the simulated light curves drawn from the mean error of the count rate in \xmm\ and \swift\ light curves.

\begin{figure}[]
\centering
\includegraphics[width=\figsize\textwidth]{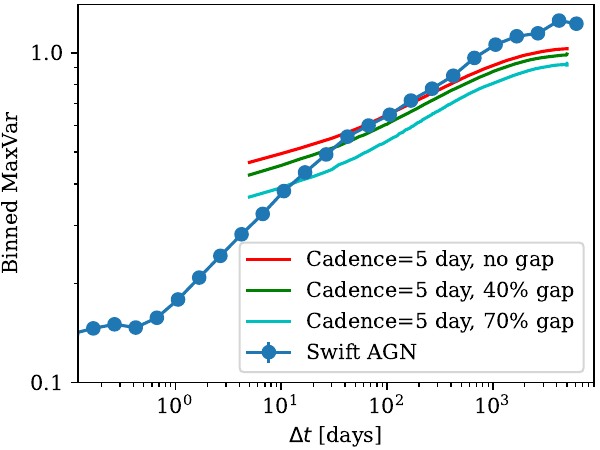}
\caption{The line connected with dots indicates the MaxVar obtained from the actual \swift\ light curves of AGNs, and the lines without dots indicate the MaxVar obtained from simulated light curves.}
\label{fig:maxvar_sim}
\end{figure}

\begin{figure}[]
\centering
\includegraphics[width=\figsize\textwidth]{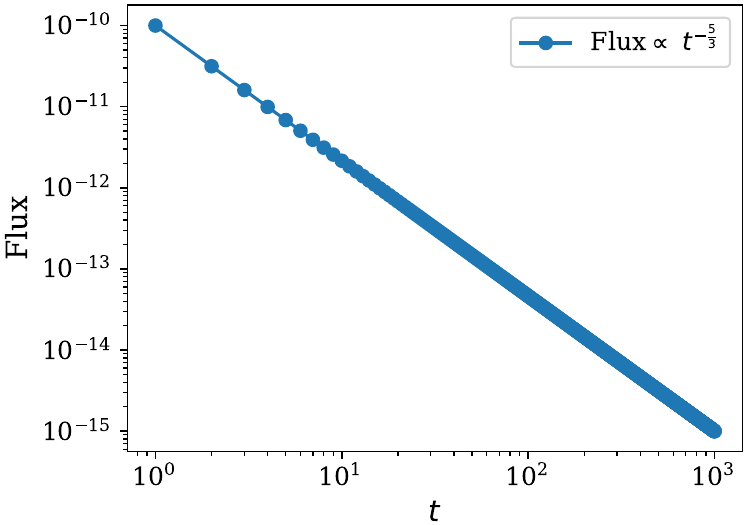}
\includegraphics[width=\figsize\textwidth]{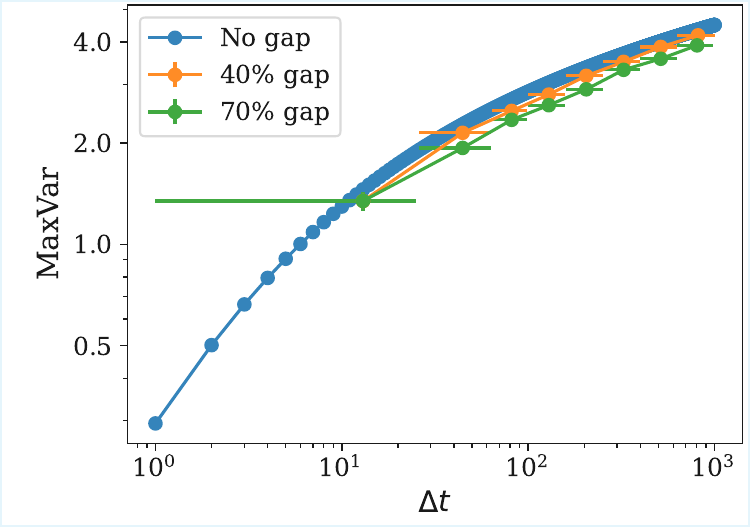}
\caption{The MaxVar computed for a single TDE following flux $\propto t^{-\frac{5}{3}}$ law. The bottom panel shows the effect of the observational gap after the first 30 days of regular monitoring. For the data set with gaps, we rebinned the MaxVar with bin size $\Delta {\rm log}(\Delta t)=0.2$ for smoothing.}
\label{fig:tde_maxvar_sim}
\end{figure}

Figure \ref{fig:maxvar_sim} shows the effect of observational gaps on the MaxVar computed from the simulated light curves. We randomly removed 40\% and 70\% of the simulated data from the light curves. It is clear that for light curves with a larger observation gap, MaxVar is decreased more for shorter $\Delta t$ than on the longer time scale. This indicates that the MaxVar computed from the \swift\ and \xmm\ light curves is underestimated at smaller time scales from the actual value, as both \xmm\ and \swift\ light curves suffered from gaps. The \xmm\ AGN light curves have more observational gaps compared to \swift, hence the MaxVar computed from the simulated \xmm\ light curve will be significantly deviated from the MaxVar obtained from sparsely sampled actual \xmm\ light curves. The gaps in the AGN light curves are visible in the Figure \ref{fig:dt_sampling}, where the tail out to large time separations becomes prominent for the light curves with the least number of data points. Furthermore, the vast majority of the light curves have fewer than 100 and 5 points in the case of \swift\ and \xmm\ AGN, respectively. In addition to that, the \xmm\ light curves have a seasonal gap of $\sim30$ days, likely associated with the solar avoidance angle. This could result in a large deviation between the MaxVar of AGN computed from the \xmm\ and \swift\ data sets as seen in Figure \ref{fig:maxvar}.

\begin{figure}[]
\centering
\includegraphics[width=\figsize\textwidth]{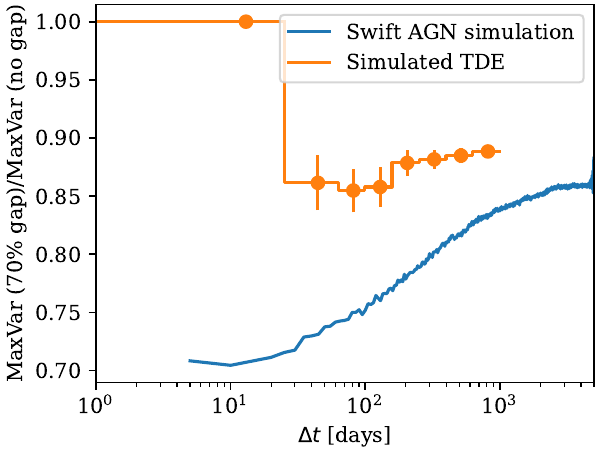}
\caption{The ratio of MaxVar obtained from light curves with 70\% and no gap for the case of AGN and TDE. The effect of gaps is stronger at shorter time scales, which can significantly reduce the MaxVar in the case of AGNs.}
\label{fig:maxvar_ratio_gap_nogap}
\end{figure}

Contrary to AGNs, the effect of observational gaps on the MaxVar of TDEs is significantly lower. This is primarily due to the fact that the long-term TDE light curves often follow a monotonic power-law decay, and the MaxVar is driven by the first few weeks of observations after the discovery of TDEs. For the TDEs in this sample, there is regular cadence monitoring for the first few weeks with \swift. The top panel of Figure \ref{fig:tde_maxvar_sim} shows the flux of an artificial TDE following $t^{-\frac{5}{3}}$ \citep{rees1988} decay with a cadence of one day. In the bottom panel of Figure \ref{fig:tde_maxvar_sim}, we show the MaxVar computed from the above TDE light curve with no gaps and 40\%, 70\% gaps after a continuous monitoring of 30 days. The deviation of the MaxVar between no gap and 40\%, 70\% gaps is between 5--12\%, depending on the time scale. Figure \ref{fig:maxvar_ratio_gap_nogap} shows the ratio of MaxVar obtained from light curves with 70\% gap and no gap for the case of AGNs and a TDE. In the case of AGNs, the gaps in the light curves significantly affect shorter time scales, which can lower the MaxVar by up to 30\% at a time scale below 10 days. However, at longer time scales, the MaxVar will be reduced by 10--15\% due to the gaps. On the other hand, in the case of a TDE with regular sampling for the first few weeks, the MaxVar is lowered by only $\sim10\%$ at both short and long time scales due to the gaps. This shows that the long-time scale gaps have a very minor effect on the MaxVar of TDEs as long as there is regular monitoring for the first few weeks after the discovery of a TDE. We note that the X-ray observations in TDEs are typically performed after the optical discovery, and missing the X-ray peak will lower the MaxVar estimation, especially for the TDEs with a steeper decline. However, in many TDEs, the X-ray peak occurs significantly later than the optical peak \citep{kajava2020,liu2022,guolo2024}, which would make it less likely for the X-ray peak to be missed.

\begin{figure}[]
\centering
\includegraphics[width=\figsize\textwidth]{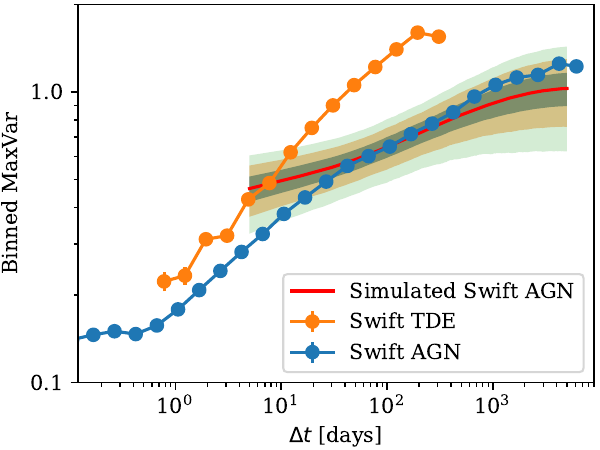}
\includegraphics[width=\figsize\textwidth]{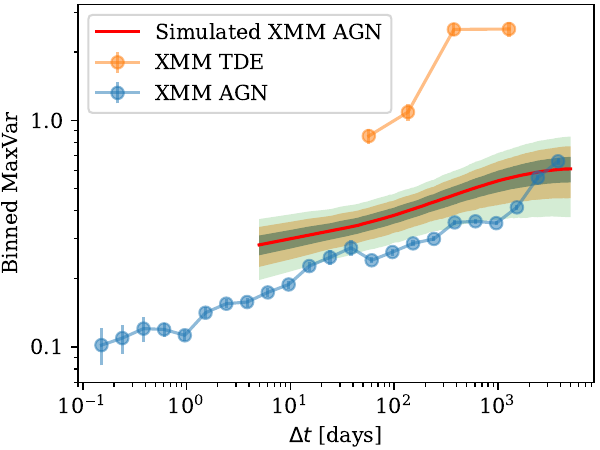}
\caption{Comparison of MaxVar for AGNs and TDEs. The red continuous line indicates the MaxVar computed from simulated light curves with a regular cadence of 5 days, and the shaded regions indicate the $1\sigma$, $2\sigma$, and $3\sigma$ confidence regions.}
\label{fig:tde_agn_maxvar}
\end{figure}

Next, we compare the MaxVar of TDEs with the MaxVar confidence intervals of AGN obtained from the simulation. The top panel of Figure \ref{fig:tde_agn_maxvar} shows the comparison for the \swift\ data set. We find that at the time scale of $\Delta t<10$ days, the \swift\ TDE MaxVar is consistent with AGN within $1\sigma$ level, therefore, it is impossible to distinguish between them from X-ray variability alone with observations of only ten days. However, at a longer time scale of around $\Delta t\gtrsim20$ days, the MaxVar of TDEs is higher than AGNs at more than $3\sigma$ confidence level. The bottom panel of Figure \ref{fig:tde_agn_maxvar} shows the comparison of MaxVar between \xmm\ TDEs and the MaxVar obtained from the simulated AGN light curves.

\subsection{Rates of TDE-like variability from AGN}
\begin{figure}[]
\centering
\includegraphics[width=\figsize\textwidth]{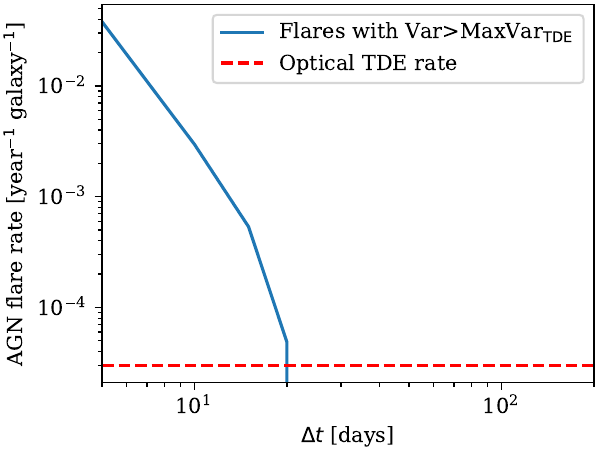}
\includegraphics[width=\figsize\textwidth]{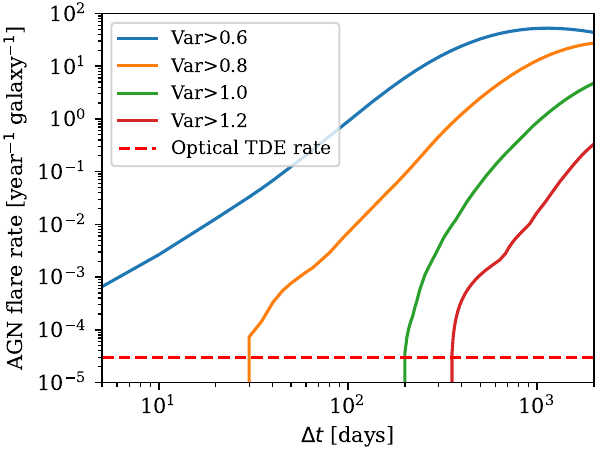}
\caption{Top panel: AGN flare rate (number of flares per unit observation time per galaxy) with flare strength above the MaxVar of TDEs as a function of $\Delta t$. The rates were obtained by counting the number of flares that have Var above the MaxVar of TDEs in the simulation of \swift\ AGN light curves. Bottom panel: AGN flare rate as a function of time with flare strength above various constant thresholds, which indicates if a transient has a Var of $>0.8$, $>1.0$, and $>1.2$ on a time scale of $<30$, $<200$, and $<300$ days, then it is highly unlikely from an AGN. The red horizontal dashed line indicates the current TDE rate from the optically selected sample. The opposite trend in the two panels is due to the fact that the Var threshold increases with $\Delta t$ in the top panel, whereas in the bottom panel, the Var thresholds are constant.}
\label{fig:flare_rate}
\end{figure}

Next, we compare the rate of AGN flare obtained from the simulation with TDE rates. In Figure \ref{fig:tde_agn_maxvar}, we have already seen that AGN flares can have a similar order of variation as TDE on a short time scale. However, looking at a longer time scale can be useful as the TDE light curves fall more rapidly than the stochastic variations in AGNs, giving rise to higher MaxVar for TDEs over AGNs. We simulated 2000 AGN light curves of 5000 days long using the model parameters obtained from fitting the SF using a power exponential model to estimate the AGN flare rate as a function of $\Delta t$. The MaxVar does not allow us to consider light curves with multiple flares, which is required for the estimation of the flare rates. Therefore, we instead consider variation (Var) as $max({\rm log}f_{\rm x})-min({\rm log}f_{\rm x})$ for all possible $\Delta t$ combinations in the simulated light curves. Then we estimate the rate as a function of $\Delta t$ by counting the number of flares that have Var above a certain threshold per time per galaxy. In our case, the length of the timeline is 5000 days, i.e., 13.7 years, and the number of mock galaxies is 2000. We compare the AGN flare rate with the TDE rate. The current estimate of the optically selected TDE rate is around $\sim3.2\times 10^{-5}$ galaxy$^{-1}$ year$^{-1}$ \citep{yao2023}. The top panel of Figure \ref{fig:flare_rate} shows the AGN flare rate that has Var above the MaxVar of TDEs as a function of $\Delta t$. As we have seen in Figure \ref{fig:tde_agn_maxvar}, at a smaller time scale, the AGN and TDE can have a similar order of variation. Therefore, at a time scale of $\Delta t<20$ days, the rate of AGN flare with a similar level of variation as TDE is extremely high. However, the AGN flare rate that has Var above the MaxVar of TDEs decreases with $\Delta t$, and at a time scale of $\Delta t>20$ days, the rate of AGN flares is much lower than the current TDE rates. This indicates that an observation time scale of $\sim20$ days or more is required to distinguish between the monotonic decline of TDEs and a flaring event from AGNs.

Using these simulated AGN light curves, we also consider the rate of AGN flares above different Var thresholds (Figure \ref{fig:flare_rate} bottom). For lower values of Var (e.g., Var of 0.6 dex), most of the transient events will be AGN flares, as the rate will be much higher than the TDE rate. However, at higher values of Var, the AGN flare rate drops below the TDE rate. For example, for a Var of 1 dex (a factor 10 flare), observing this variation over $\sim200$ days has an equal chance of being an AGN flare or a TDE, but observing this level of variation over $<200$ days is more likely to be a TDE. More generally, if a transient is observed with a variation of $>0.8$, $>1.0$, and $>1.2$ dex on a time scale of $<30$, $<200$, and $<300$ days, respectively, then it is highly unlikely to be an AGN flare.

\section{Discussion}\label{sec:discussion}

\subsection{Turnover in the AGN SF}
The goal of this work is to utilize the X-ray variability for the characterization of TDE emission and flares from AGN. To this aim, we estimate the average variability of the AGN sample using ensemble-averaged SF. We fitted the SF using three models: a simple power-law model, a broken power-law model, and a power exponential model. X-ray variability is typically analyzed in the frequency domain $f$ using power spectral density (PSD), and the PSD slope is related to the SF slope by $\alpha=1+2\gamma$, where ${\rm PSD}(f)\propto f^{-\alpha}$. For the power-law model fits to our data, the SF slope corresponds to the PSD with a power-law index $\alpha=1.22\pm0.02$, and $1.27\pm0.08$ for the \xmm\ and \swift\ AGNs, respectively. For both a simple power-law and a broken power-law model, the X-ray SF slope is shallower than the estimates from optical studies. The optical studies of AGN suggest that below the flattening time scale, the power-law index of SF is $\sim0.5$ \citep{kozlowski2016}, indicating that optical SF is steeper than X-ray, hence the X-ray has more variability power over optical light curves at shorter time scales. This has also been confirmed through the PSD analysis of {\it Kepler} data that shows the optical PSDs of AGNs have a power-law index of 2.6--3.3, which is much steeper than X-ray PSDs \citep{mushotzky2011,smith2018}. 

The X-ray PSD of AGNs shows multiple breaks. Typically, X-ray PSD of AGN is known to have $\alpha\sim2$ and $\alpha\sim1$ above and below a high-frequency break, respectively \citep{markowitz2003,mchardy2006}. \citet{mchardy2006} suggests the high-frequency break time scale is likely associated with the inner edge of the accretion disk. However, there should also be another low-frequency break (break frequency associated with the damping time scale $\tau$), below which the PSD flattens simply because the variability amplitude cannot increase forever. Contrary to optical studies, it is extremely difficult to detect the damping time scale in X-rays for individual AGNs above which the variability amplitude becomes constant. This is in part due to the lack of long X-ray monitoring of individual objects, and estimating the low-frequency break from PSD requires well-sampled light curves with no gaps. However, this could be done for AGNs with smaller-mass BHs, for which the damping time scale is expected to be lower. So far, flattening of the X-ray PSD has been seen in three AGNs: Ark 564 \citep{mchardy2007}, NGC 3783 \citep{markowitz2003}, and NGC 5506 \citep{mchardy1988}. 

We find evidence of flattening in the ensemble average SF of AGNs that can be fitted by a power exponential model. The damping time scale obtained by fitting the power exponential model is $\tau=950\pm300$ days. The damping time scale is slightly higher than the values found in optical studies. Optical variability studies of AGN using a damped random walk model suggest a damping time scale of $\tau\sim10^2$--$10^3$ days depending on the BH mass \citep{kelly2009,kozlowski2010,macleod2010}, with a median value of $\tau=0.97\pm0.46$ years \citep{kozlowski2016}. Furthermore, the damping time scale $\tau$ obtained from modeling of optical light curves of AGN indicates a linear correlation with $\tau$ and BH mass \citep{kelly2009,kozlowski2016,burke2021}. Despite the differences in the power-law index between X-ray and optical studies, the theoretical model of AGN variability predicts that the damping time scale for optical and X-ray light curves should be similar \citep{lyubarskii1997}. According to the mass accretion rate fluctuation model for the origin of AGN variability, the variation at the outer disk is propagated inward, hence modulation of optical and infrared is imprinted in the X-ray light curves, which should give a similar damping time scale across all wavebands \citep{lyubarskii1997,hagen2024}. Furthermore, according to the mass accretion rate fluctuation model, the high-energy photons should lag behind low-energy photons when looking at the long-time scale variations that have been seen in Fairall 9 \citep{hernandez-santisteban2020}.

\subsection{MaxVar of AGN flares and eROSITA TDEs}

\begin{figure}[]
\centering
\includegraphics[width=\figsize\textwidth]{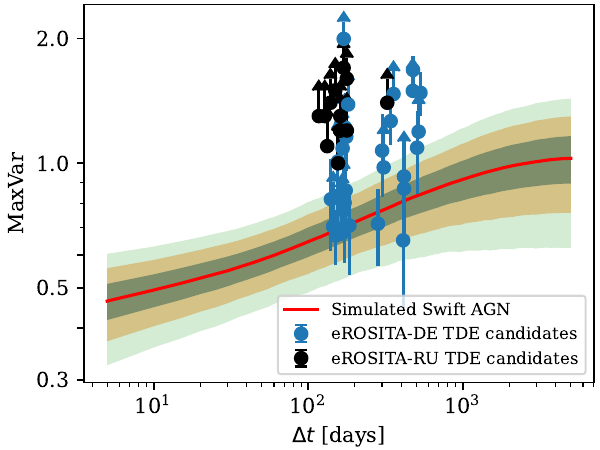}
\caption{The comparison of MaxVar obtained from simulated AGN light curves with eROSITA TDE candidates.}
\label{fig:compare_erosita}
\end{figure}

Next, we compare the MaxVar of eROSITA TDE candidates with the expected maximum variation from AGN flares. So far, there are four all-sky scans completed by eROSITA, with subsequent scans separated by six months. The majority of the eROSITA TDE candidates have detections only during the peak and upper limits in the rest of the pointings. Therefore, in Figure \ref{fig:compare_erosita}, the vast majority of MaxVar measurements of eROSITA TDEs are a lower limit. The German and Russian sides of eROSITA scans reported detections of 31 and 13 TDE candidates, respectively \citep{szanov2021,grotova2025}. The selection criteria for TDE candidates in the eROSITA-RU scan are that non-detection in the eROSITA first scan, and the flux level in the second eROSITA scan exceeds at least a tenfold of the first scan upper limit \citep{szanov2021}. Hence, all the eROSITA-RU TDE candidates have MaxVar above 1 dex, and they are well separated from the MaxVar of AGNs. On the other hand, many of the eROSITA-DE TDE candidates have a MaxVar consistent with AGNs. This is due to the fact that eROSITA-DE TDE candidates were selected primarily based on decline or rise-like shape in the light curve and X-ray spectral photon index $\Gamma\geq 3$, but with no threshold on flux differences \citep{grotova2025}.

It is unclear how effective the X-ray spectral slope is in separating TDEs from AGNs. Some studies of TDEs have found soft ($\Gamma\geq 3$) X-ray spectral slopes, with clear separation from AGN \citep{guolo2024}. On the other hand, there are at least three known jetted TDEs \citep[SwJ1644, SwJ2058, and AT2022cmc;][]{bloom2011,burrows2011,cenko2012,andreoni2022} that have significantly hard emission, which is likely originating from the base of the jets. \citet{auchettl2018} compared the X-ray hardness ratio for four TDEs and four extremely variable AGNs, finding significant overlap. Furthermore, some TDEs show the emergence of hard X-ray emission at late times, suggesting that a variety of X-ray emission properties can be seen in TDEs \citep{guolo2024}. Very soft X-ray spectra can also be seen in non-TDE sources; for example, AGN during peak flaring events exhibit a similar soft X-ray spectrum compared to the pre- or post-flare state \citep{saha2023,krishnan2024}. Thus, very soft X-ray spectra ($\Gamma\geq 3$) may not be sufficient to uniquely identify TDEs. The MaxVar comparison suggests many of the eROSITA-DE TDE candidates may not be TDEs, and those with the MaxVar consistent with AGN should be followed up with deeper X-ray observations to confirm whether they follow a steep power-law-like decay expected from TDE or have a higher MaxVar lower limit than AGNs.

\section{Conclusions}\label{sec:conclusions}
We cross-matched samples of AGNs and TDEs from the literature with 4XMM-DR14s and LSXPS X-ray catalogs of \xmm\ and \swift, respectively. We computed the X-ray variability of AGNs using SF and later used the best-fit SF model to simulate the light curves with regular cadence. Furthermore, we computed the MaxVar of the simulated light curve and compared it to the TDEs. The main conclusions of this study are as follows.
\begin{itemize}
    \item The variability of AGNs can be largely described by SF with a power-law index $\gamma\sim0.11-0.14$. Furthermore, the ensemble average SF of AGNs shows flattening at longer $\Delta t$. Fitting the SF using a power exponential model indicates a damping time scale $\tau=989\pm256$ and $930\pm328$ days for \xmm\ and \swift\, data sets, respectively.
    
    \item We find that at a smaller time scale $\Delta t<20$ days, AGNs can have a similar order of variation as TDEs. Therefore, a longer observation of $\sim20$ days or more will be required for classification between AGN and TDEs. The rate of AGN flares that have Var above the MaxVar of TDE decreases with $\Delta t$. Furthermore, if a transient has a Var of $>0.8$, $>1.0$, and $>1.2$ on a time scale of $<30$, $<200$, and $<300$ days, then it is highly unlikely to be an AGN. 

    \item The MaxVar comparison of eROSITA TDE candidates suggests many of the eROSITA-DE TDE candidates are consistent with AGNs, suggesting some of them may not be TDEs; therefore, a further follow-up observation is required for their secure classification. 
\end{itemize}

\begin{acknowledgments}
SM and KDF acknowledge support from the grant NASA ADAP 80NSSC24K0666. We thank the anonymous referee for useful feedback, which has improved this manuscript.
SM thanks Szymon Kozlowski for help with the light curve simulations using a power exponential model of SF. 
\end{acknowledgments}

%

\vspace{5mm}
\facilities{\swift, \xmm\ }


\software{python \citep{vanrossum2009}, jupyter \citep{kluyver2016}, astropy \citep{2013A&A...558A..33A,2018AJ....156..123A}, numpy \citep{vanderwalt2011,harris2020}, matplotlib \citep{hunter2007}}

\bibliography{sample631}{}
\bibliographystyle{aasjournal}



\end{document}